\documentclass[twocolumn,showpacs,preprintnumbers,amsmath,amssymb]{revtex4}

\usepackage{graphicx}
\usepackage{dcolumn}
\usepackage{bm}

\begin{document}

\title{The number of capillary bridges in a wet granular medium}

\author{Dimitrios Geromichalos}
    \email{Dimitrios.Geromichalos@physik.uni-ulm.de}

\author{Mika M. Kohonen}

\author{Mario Scheel}

\author{Stephan Herminghaus}

\affiliation{%
    Applied Physics Lab., Ulm University, Ulm, D-89069,
    Germany
}

\date{\today}

\begin{abstract}
We observed the appearance of capillary bridges in a granular
medium consisting of glass beads after adding small amounts of
liquid. We found the initial bridge formation depending on the
bead roughness. Furthermore we obtained a statistics for the
average number of bridges for randomly packed beads in dependence
of the liquid content and were able to find an explanation
therefore based on recent models and former experimental data.
\end{abstract}


\maketitle

If one adds a wetting liquid to a granulate, there occur capillary
bridges which exert forces on the particles. These forces are the
main reason for the strongly changed  mechanical properties of wet
granular matter compared to the dry case \cite{mik98,
israelachvili, horn97, hal98, boq98, fra99, ger03}. Despite the
notable advance in the comprehension of the dynamical properties
of dry granular materials \cite{ristow, jae96, kad99}, the
physical mechanisms, which are the cause for the character of wet
systems, remain mostly unknown.

Since the formation of the liquid bridges depends apparently
strongly on the microscopic geometry of a granulate, an adequate
description of the packing is essential. The random close packing
of beads, which is the simplest random packing and which occured
in our experiments presented in this paper, is an interesting and
challenging topic in itself and has been much studied in the past
\cite{sco60, mas68, pet01}. It was the goal of our experiments to
show how the formation of capillary bridges, which are essential
for the mechanics of a wet granular matter \cite{ger03, boq98,
fen00, sim93, mik98}, depends on the geometrical particle
distribution, which was subject of earlier studies
\cite{mas68,pet01}.

In order to observe and count the single bridges we used an index
matching technique \cite{jai01}. At this we put glass beads with
diameters of $375 \mu m$ or $555 \mu m$, which had a refraction
index of 1.51, into a mixture of toluene and diiodomethane with
the same refractive index, so that the granular matter became
quite transparent and we were able to zoom with the microscope
through the glass bead layers and observe the bridges. In order to
achieve this index matching the volume share of the toluene was
$88.1 \%$ and of the diiodomethane $11.9 \%$, respectively. This
liquid mixture took over the role of "air" in a "normal" wet
granular matter. As the wetting liquid, i.e. as liquid which forms
the capillary bridges, we added small amounts of water with
fluorescein.

The granular matter was put into a cuvette ($0.95 \times 0.95
\times 4 cm$, filling height: $2.5 - 3 cm$) and was shaken in
small horizontal circles with an amplitude of $5 mm$ and with a
frequency of roughly $30 Hz$ for some minutes, until the "wetting
liquid", clearly visible due to fluorescein, appeared
homogeneously distributed. The diameter of the glass beads used
was $375 \mu m$ and $555 \mu m$. In order to prevent
crystallization, the small beads were chosen to be slightly
polydisperse: the spread in bead size was $10 \%$ for the glass
beads with a diameter of $375 \mu m$ and $1 \%$ \footnote{Although
a spread of only $1 \%$ can lead to a crystallization of the
granulate, it has not always to be the case. In our experiments
presented in this paper we observed no crystallization.} for the
glass beads with a diameter of $555 \mu m$. After shaking, the
volume fraction of the glass was $0.57$, i.e. we found the
granulate packed less densely as it would be the case for the
loose random packing of $0.60$ for the dry case \cite{sco60}. We
could also reach a random dense packing of $0.62$, which is the
other extreme of the random packing of glass beads and is similar
to a volume fraction of the glass of $0.64 =: \rho_{rdp}$
\cite{sco60}, by tapping the sample several times.

After adding very small amounts of liquid, there was no formation
of capillary bridges. The reason for this behavior is that the
liquid was first trapped on the surface, as it can be seen in Fig.
1a. Only at a water content of $0.07 \%$, bridges on almost every
bead contact were formed (see below). Fig. 1b illustrates the case
for fully formed bridges.

\begin{figure}[h]
 \includegraphics[width = 7.5cm]{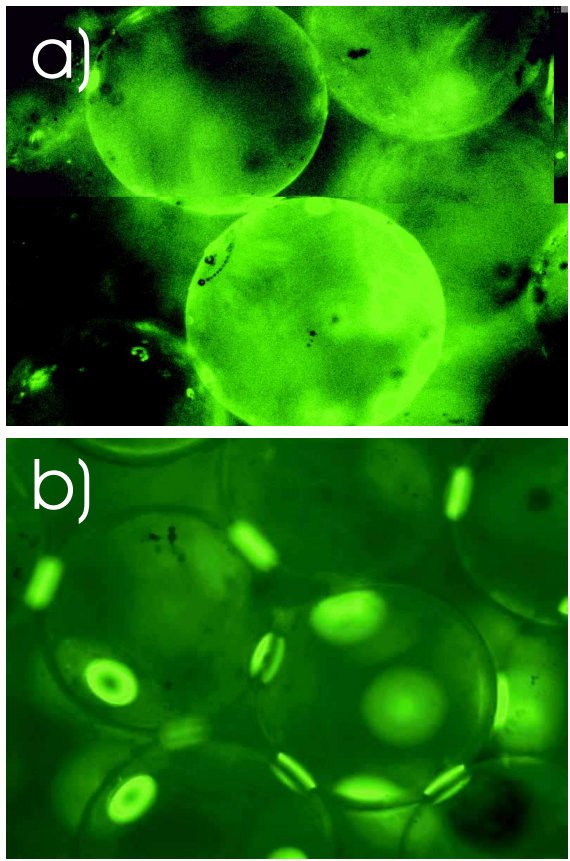}
 \caption{Fluorescence microscope pictures for different water contents $W$; glass
 bead diameter: $375 \mu m$ \newline
 1a) $W = 0.055 \%$, no real bridges have formed yet \newline
 1b) $W = 0.2 \%$, capillary bridges between the beads are clearly visible}
\end{figure}

It is well known that bridges need a certain amount of liquid
content $W$ to form properly, since the liquid is at first bound
on the bead surfaces due to the roughness \cite{hal98, wil00,
horn97}. The connection between the roughness amplitude $\delta$
and the saturation bridge volume\footnote{The reason for calling
$w_b$ saturation bridge volume comes from the Halsey-Levine
theory. According to this modell, the force exerted by bridges
with a volume $w>w_b$ does no longer increase significantly with
$w$. \cite{hal98}} is given by $w_b \approx 2 R \delta^2$, where
$R$ is the bead radius, and $w_b$ the volume of a bridge at the
water content of saturation $W_b$ \cite{horn97}. Our measurements
\cite{koh03} suggest that the relationship between the water
content $W$ and the volume of a single bridge $w$ is generally
given by $w = \alpha r^3 W$, where $\alpha \approx 0.25$.
\cite{ger03}

From $W_b \approx 0.07 \%$ we get $\delta \sim 500nm$, which is
similar to the peak-to-peak roughness we found from the inspection
of the beads by atomic force microscopy.

We obtained the average number of bridges per bead by zooming with
the microscope through the sample and counting the bridges. We
carried this out for different liquid contents for the random
loose packing which was in our case 0.57 as well as for the random
dense packing of 0.62. In order to reach better statistics we
counted the bridges of 40 beads for each data point for the dense
packing and 20 beads for the loose packing and calculated the
average bridge number $N$. We took the standard deviation as error
for $N$. Since we counted only bridges from beads of the second to
the sixth layer, an estimation of the influence of the sample
boundary is necessary. Therefore we measured $N$ for different
layers $L$. We found that $N(L) = const$ for $3 \leq L \leq 6$,
whereas for $L \leq 2$ small variations were observed. We believe
therefore that we can exclude any effect of the sample boundary on
$N$.

\begin{figure}[h]
 \includegraphics[width = 8.5cm]{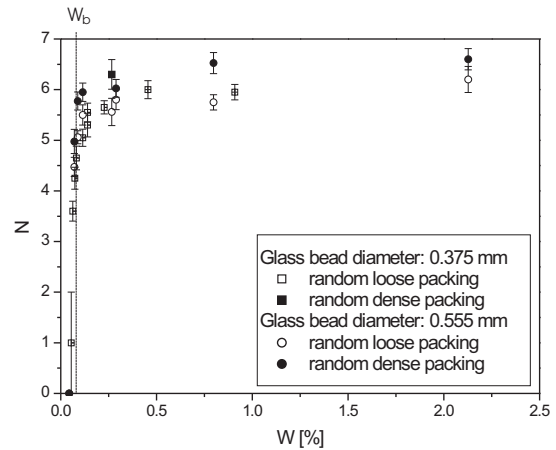}
 \caption{Average number of capillary bridges $N$ in dependence of
 the water content $W$}
\end{figure}

The dependence of $N$ on the water content is shown in Fig. 2. $N$
is equal to zero for very small $W$. At a water content slightly
smaller than $W_b$ it goes up very fast, then the curve becomes
flatter. It can be seen further, that $N(W)$ for the random dense
packing is about $10 \%$ larger than $N(W)$ for the random loose
packing.

As it was mentioned earlier, the number of capillary bridges is
influenced strongly by the microscopic geometrical properties of
the granular medium. A model which deals with the number of
contacts of randomly packed glass beads is the caging model
\cite{pet01}. According to this model one needs $4.79 =: N_{dc}$
bead contacts on average in order to pin a bead with other beads.
We assumed that this model provides at least roughly an
appropriate explanation for our experiments and have defined $W_b$
accordingly. Consequently $W_b$ is the water content at which a
capillary bridge exists then and only then, if two beads touch,
i.e. if their surface distance $d$ is zero. The smaller non-zero
values at $W < W_b$ can be explained as follows. The water film on
the beads varies in thickness at different areas of the bead
surface. Therefore it is possible at certain water contents that
bridge formation occurs only at a fraction of the bead contacts.
On the other hand, for $W>W_b$ there can also occur slightly
"streched" bridges between beads which do not touch, but whitch
are close enough to each other (see below).

Our assumption that the caging model is appropriate for our case
is confirmed further if we compare the measured distribution of
the capillary bridges at water contents $W \approx W_b$ for the
random loose packing to the distribution obtained by a simulation
\cite{pet01} based on the caging model. As it can be seen in Fig.
3, the measured curve is very similar to the simulated one.
Particularly the measured curve does not exceed the value of 8,
i.e. it fulfills a necessary condition for the validity of the
caging model \cite{pet01}. At higher water contents the histogram
curve is shifted to the right, but we did never see a glass bead
with more than 9 capillary bridges.\footnote{At very high liquid
contents ($W \geq 3 \%$) there form liquid clusters between the
beads, so that the bridge concept is no longer valid.}.

\begin{figure}[h]
 \includegraphics[width = 8.5cm]{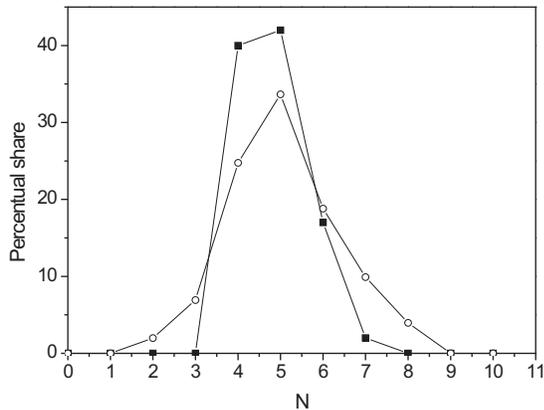}
 \caption{Distribution of the number of bridges: \newline
 measured ($W = 0.15 \%$, random loose packing; open symbols) \newline
 simulated \cite{pet01} (closed symbols)}
\end{figure}

We will compare now our measurements concerning the dependence of
the number of capillary bridges on $W$ to the measured distance
distribution of beads in a random close packing \cite{mas68}. For
$W \geq W_b$ there exists at least at every bead contact (i.e. at
every point with $d=0$, where $d$ is the surface distance of the
beads) a capillary bridge: $N \geq N_c$. $N_c$ is here the number
of real contacts per bead. According to the caging model we have
$N_{dc} = 4.79$. \cite{pet01} Bridges can also exist between beads
which do not touch, but which are only sufficient close to each
other. Due to the hysteretic nature of bridge formation
\cite{lothar02}, it is reasonable to suppose that such bridges can
form only between beads which collided in the past and then moved
away from each other. For sufficient strong shaking this should be
statistically the case for a certain part $A^*$ of the beads, i.e.
\[
N(W) = N_c + A^* N_b(W)
\]
Here $N_b$ is the number of nearest neighbors, which do not touch
the bead, but whose distance $d$ is smaller than the bridge
rupture distance $d_p$. The total number of neighbors in between a
distance $d_p$ is
\[
N_t(W) = N_c + N_b(W) = N_c + \frac{N(W)-N_c}{A^*}
\]
Herewith we get
\begin{equation}
\label{bridgenum:eq} N(W) = A^* N_t(W) + (1 - A^*) N_c
\end{equation}
We need now to correlate the water content $W$ to rupture distance
$d_p$. According to \cite{willet00} this correlation is
approximately given by
\[
d_p = w^{\frac{1}{3}} + \frac{1}{10} \frac{w^{\frac{2}{3}}}{R}
\]
where $R$ is the bead radius and $w$ the bridge volume. In
accordance to our own measurements \footnote{M. M. Kohonen, D.
Geromichalos, and S. Herminghaus; to be published}, we have $w =
\alpha R^3 W$ with $\alpha \approx \frac{1}{4}$. Therefore
\[
d_p = R \left(  (\alpha W)^{\frac{1}{3}} + \frac{1}{10} (\alpha W
)^{\frac{2}{3}} \right)
\]
Now it is possible to calculate $N$ from the measurements of $N_t$
\cite{mas68}, plot it over $d_p$ and compare it with the directly
measured number of capillary bridges. Before doing this we defined
$W^* := W - W_b$, i.e. the water content above $W_b$ which is
located in the bridges. The reason therefore is that the glass
bead surface absorbs a certain amount of liquid $W_b$ before
bridge formation starts. Accordingly we defined $d_p^* :=
d_p(W^*)$ as corresponding rupture distance.

\begin{figure}[h]
 \includegraphics[width = 8.5cm]{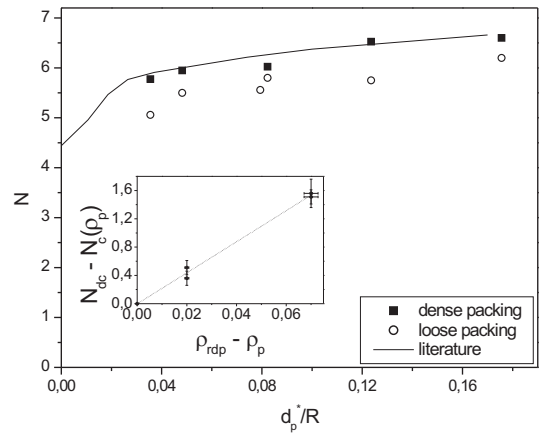}
 \caption{Comparison of the measured number of bridges $N$ with random loose and random dense packings to
 the $N$ which were calculated from the literature values \cite{mas68} according Eq. \ref{bridgenum:eq} \newline
 Inset: Number of bead contacts for different packing densities.}
\end{figure}

In order to get a quantitative expression the values of $A^*$ one
has to consider the mechanism of bridge formation. A bridge
between two non-touching beads can only exist if their distance is
smaller than $d_p$ if this beads touched each other in the past.
Therefore one would expect $A^* = \frac{1}{2}$ for a
one-dimensional system with statistically moving particles. In a
real granulate the situation is more complicated. It is possible
that particles move against each other by a distance smaller than
$d_p$ without ever touching, which leads to a decrease of the
value of $A^*$.

In order to obtain an estimation for $A^*$ we considered a single
bead with radius $R$ in the point of origin. Then we assumed other
beads with radius $R$ moving by at an impact parameter $p^*$ which
was chosen at random. It is evident that bridges can only form if
$0 \leq p^* \leq 2R$, i.e. a bead collision is a necessary
condition for bridge formation. If the moving bead moves along the
x-axis, the distance between the surfaces is given by $d =
\sqrt{{x^*}^2 + {p^*}^2}-2R$. At a certain point of time a bridge
can only exist if $d \leq d_p^*$. A further condition for the
existence of a bead is that the collision took back in the past
which means that only half of the beads with $0 \leq p^* \leq 2R$
and $d \leq d_p^*$ will have a bridge. The arc length of a bead
which has a bridge is given by
\[
B_1(p^*) = \int_{x_1}^{x_2} \sqrt{1 + \frac{x^2}{x^2 + {p^*}^2}}
\mbox{d}x
\]
with $x_1 ={\sqrt{(2R)^2 - {p^*}^2}}$ and $x_2 = {\sqrt{(d_p^* +
2R)^2 - {p^*}^2}}$. The arc leangth of a (bridgeless) bead with
$2R < p^* \leq 2R + d_p^*$ and $d^* \leq d_p^*$ is analogous
\[
B_1(p^*) = 2 \int_{0}^{x_2} \sqrt{1 + \frac{x^2}{x^2 + {p^*}^2}}
\mbox{d}x
\]
With $I_1 := 2 \pi p^* \int_0^{2R} B_1 ({p^*}) \mbox{d}p^*$ and
$I_2 := 2 \pi p^* \int_{2R}^{2R+d_p^*} B_2 ({p^*}) \mbox{d}p^*$
one finally gets
\[
A^*(d_p^*) = \frac{I_1}{2 I_1 + I_2}
\]
$A^*(d_p^*)$ can be approximated very well as follows:
\[
\frac{1}{A(d_p^*/R)} \approx 1.447 \frac{d_p^*}{R} + 2
\]
Fig. 4 shows the plots of our measured $N$ over $d_p^*$ for the
beads with a diameter of $555 \mu m$ and for the random loose
packing and the random dense packing. The drawn curve represents
the values of $N$ for the dense packing which were calculated from
the literature values of $N_t$ \cite{mas68} using Eq.
\ref{bridgenum:eq}. It is clearly visible that our calculated
values lie near to those of the literature for $N_c = 3.3$ for the
loose and $N_c = 4.3$ for the dense packing. $N_c$ differs from
the $N_c$ of the caging model \cite{pet01} significantly for the
loose packing, while the difference in the dense packing is quite
small. Hence the assumption that $N_c$ depends directly on the
packing density $\rho_p$ is self-evident. The measured
$N_c(\rho_p)$ as well as the theory values are in agreement with
the linear curve $N_{dc} - N_c (\rho_p) = 22 (\rho_{rdp} -
\rho_p)$ (see Fig. 4 (inset)). This plot displays the $N_c$ of the
loose as well as the dense packing for the used beads with a
diameter of $555$ and $375 \mu m$. The point of origin corresponds
to the $N_{dc}$ of the caging model for the random dense packing
$\rho_{rdp}$. Concluding the value of the number of contacts
according to the caging modell \cite{pet01} $N_{dc}$ is given by
the linear extrapolation of the $N_c (\rho_p)$ obtained from our
measurements: $N_{dc} = N_c (\rho_{rdp})$.

The authors thank the German Science Foundation for financial
support within the Priority Program `Wetting and Structure
Formation at Interfaces'. MMK acknowledges the Alexander von
Humboldt Foundation for generous funding.

\bibliography{apssamp}

\end{document}